\documentclass[12pt]{article}

\usepackage{scicite}

\usepackage{times}
\usepackage{url}
\usepackage{hyperref}

\newenvironment{sciabstract}{%
\begin{quote} \bf}
{\end{quote}}

\newcounter{lastnote}
\newenvironment{scilastnote}{%
\setcounter{lastnote}{\value{enumiv}}%
\addtocounter{lastnote}{+1}%
\begin{list}%
{\arabic{lastnote}.}
{\setlength{\leftmargin}{.22in}}
{\setlength{\labelsep}{.5em}}}
{\end{list}}

\title{Gas Disks to Gas Giants:  Simulating the Birth of Planetary Systems}

\author
{Edward W. Thommes,$^{1,2\ast}$ Soko Matsumura,$^{2}$ Frederic A. Rasio$^{2}$\\
\\
\normalsize{$^{1}$University of Guelph, Guelph, ON, Canada}\\
\normalsize{$^{2}$Northwestern University, Evanston, IL, USA}\\
\\
\normalsize{$^\ast$To whom correspondence should be addressed; E-mail:  thommes@northwestern.edu.}
}

\date{}

\usepackage{epsfig}
\usepackage{url}

\begin{document}


\baselineskip24pt

\long\def\symbolfootnote[#1]#2{\begingroup%
\def\thefootnote{\fnsymbol{footnote}}\footnote[#1]{#2}\endgroup}

\newcommand{\apj}{Astrophys. J.}
\newcommand{\apjl}{Astrophys. J.}
\newcommand{\aj}{Astron. J.}
\newcommand{\mnras}{Mon. Not. R. Astron. Soc.}
\newcommand{\nat}{Nature}
\newcommand{\aap}{Astron. Astrophys.}
\newcommand{\araa}{Annu. Rev. Astron. Astrophys.}

\maketitle

\begin{sciabstract}

The ensemble of now more than 250 discovered planetary systems 
displays a wide range of masses, orbits and in multiple systems, dynamical interactions.  These represent the endpoint of a complex sequence of events, wherein an entire protostellar disk converts itself into a small number of planetary bodies.  Here we present self-consistent numerical simulations of this process, which produce results in agreement with some of the key trends observed 
in the properties of the exoplanets.
Analogues to our own Solar System do not appear to be common,
originating from disks near the boundary between barren and (giant) planet-forming.
\end{sciabstract}

Gas giant planets form within 1-10 million years (Myrs), during the time that their parent star possesses a gas disk\cite{2001ApJ...553L.153H}.
Two- and three-dimensional hydrodynamic simulations of planets embedded in protostellar disks are, due to their computational cost, limited to following a maximum of $10^3-10^4$ orbits; even for a planet as far out as Jupiter (orbital period 11 years), this amounts to at most a tenth of a protostellar disk's total lifetime.  For this reason, longer-timescale simulations usually model the time after the protostellar gas disk has dissipated, allowing the problem to be tackled with pure gravitational $N$-body methods.  Recent work in this area has been successful at reproducing the observed exoplanet eccentricity distribution, the key requirement being simply that planets must begin close enough to each other to render systems dynamically active from the outset\cite{2007astro.ph..3160J,2007astro.ph..3166C}.
Although these results provide support for planet-planet interaction as the main agent behind the eccentricities, they do not address the question of how, or even if, planet formation produces the requisite crowded systems.  Furthermore, many of the observed cases appear to have kept a memory of their early dynamical evolution:  About a quarter of discovered systems contain planet pairs locked into dynamical mean-motion resonances\cite{2007prpl.conf..685U},
likely the product of early differential migration\cite{2002ApJ...567..596L} via planet-disk interaction\cite{1980ApJ...241..425G,1986ApJ...309..846L,1997Icar..126..261W}.
Thus, to achieve a more complete understanding of the planet formation process, it is essential to bridge the disk and post-disk era.  Most models of this regime
have been either semi-analytic treatments, which do not consider planet-planet interactions\cite{2004ApJ...604..388I,2005A&A...434..343A,2006ApJ...652L.133C,2008ApJ...673..487I}, or $N$-body simulations with parameterizations of disk effects\cite{2003Icar..163..290A,2008arXiv0801.1926L}.
Here, we perform simulations using a hybrid numerical scheme\cite{2005ApJ...626.1033T}, that combines the N-body code SyMBA (Symplectic Massive Body Algorithm)\cite{1998AJ....116.2067D} with a
one-dimensional disk model that self-consistently interacts with, and accretes onto, the embedded planets.  We began our simulations with the appearance of the first small protoplanets ($10^{-3}\,M_{\rm E}$ where $M_{\rm E}$ is the Earth's mass) in a gas disk and ran them for up to half a billion years (Gyrs).  Gas giants are assumed to form via core accretion (see Supporting Material for a more detailed description of the code).

The fundamental question we sought to address is how the properties of a mature planetary system map to those of its birth disk.  To this end, we performed simulations covering a range of disk parameters;  Fig. 1 depicts one representative example, which illustrates some of the key effects at work in the planet formation process (see Supporting Material for an animated version,
as well as additional examples).
In this simulation, protoplanet growth initially proceeded in an orderly manner; the dissipational effect of the gas disk dominated over the bodies' mutual perturbations, so that all orbits were kept nearly circular.  Simultaneously, interaction with the disk also caused protoplanets to undergo orbital decay, referred to as ``type I migration"\cite{1997Icar..126..261W}.  This effect will in general deposit some protoplanets at the inner disk edge (or onto the central star) before they have a chance to become gas giants \cite{2005A&A...434..343A}, but since migration speed scales with migrator mass and gas disk density, those protoplanets that grow slowly enough relative to the disk's depletion time avoid this fate\cite{2007ApJ...656L..25T,2008ApJ...673..487I}.     All protoplanets in the simulation (Fig. 1) fell into the latter category (see Supporting Material for counterexamples), although the orbit of the innermost one, originating at 2 AU, shrank to
$1/5$ its original radius by 1 Myr.  Around this time, the first gas giant formed, scattering many of the neighboring protoplanets to high eccentricities as its mass rapidly grew.  Its perturbation on the surrounding gas also became significant, and a deep annular gap opened around its orbit.  With gas unable to readily cross its orbit, the planet was locked into the disk and carried along by the latter's accretion flow, in what is termed ``type II migration"\cite{1997Icar..126..261W}.
At 1.2 Myrs, the gap become a hole as the inner disk material drained away faster than the planet migrated; however the hole rapidly closed in again, depositing the planet at the inner disk edge at 1.24 Myrs.
By this time two more gas giants began to form.  Each in turn opened a hole in the disk, in the process cutting off its inner neighbor from the gas, so that ultimately only the outermost planet was directly interacting with---and accreting from---the disk.  The outer planet migrated toward the middle one, and at 1.66 Myrs, the two become locked into a 3:1 mean-motion resonance\cite{1999..Murray..Dermott..book}
with each other, mutually increasing their eccentricities as they moved inward together.  Migration and further planet accretion stalled as the disk density became low; the disk disappeared altogether a little after 4 Myrs, leaving behind a system consisting of three gas giants---the outer two still in resonance---plus an outer Neptune-class planet.  This system remained stable for the rest of the simulation, which ended after 0.5 Gyr.

We performed a set of 100 simulations over a range of parameters.  The properties of protoplanetary disks are only weakly constrained; 
two key parameters are initial gas disk mass $M_{\rm disk}$, and disk viscosity $\nu$, which determines how rapidly accretion onto the central star removes the bulk of the gas [photoevaporation may remove the last of the gas\cite{2000prpl.conf..401H}].  Observations combined with modelling\cite{1998ApJ...495..385H} suggest that, roughly, $10^{-2} M_{\rm Sun} < M_{\rm disk} <  10^{-1} M_{\rm Sun}$ and $10^{-3} < \alpha < 10^{-2}$,
using the common parameterization\cite{1973A&A....24..337S} $\nu \equiv \alpha c_s H$,
with $c_s$ the gas sound speed and $H$ the disk scale height.
We used this range of parameters for our simulations; the resulting planetary systems (Fig. 2)  were each the product of a complex interplay between planet-disk and planet-planet effects, thus individual outcomes were highly stochastic.  Nevertheless, clear trends with $M_{\rm disk}$ and $\alpha$  are visible:  In one extreme, low disk mass combined with high viscosity resulted in systems that produced no gas giants at all.  In the other extreme, high disk mass combined with low viscosity results in the production of numerous gas giants; most underwent significant inward migration, and many acquired large eccentricities.

We can understand these results in terms of two fundamental timescales in a planet-forming disk:  One is the gas disk depletion time $\tau_{\rm disk}$,
the other is the time to form the first gas giant, $\tau_{\rm giant}$.  As shown in Fig. 3, we expect an initial burst of planet formation that spreads out from a particular radius, typically comparable to the Jupiter-Saturn region of our own Solar System (5-10 AU).  As time passes, the delay between the birth of successive giants becomes longer and longer.  Thus, the crucial factor determining how a given system's formation history will play out is the time during the gas disk's lifetime that this burst occurs.  In cases with $\tau_{\rm giant} > \tau_{\rm disk}$ (Fig. 3, lower right), the gas is removed before any gas giant has a chance to form, leaving behind systems consisting solely of rocky-icy bodies.  In cases with $\tau_{\rm giant} < \tau_{\rm disk}$ (upper-left region of Fig. 2) planets are born into a substantial gas disk, and such systems generally produced a number of gas giants that migrated inward significantly.  Planet-bearing systems can be further classified into ``planet-dominated" ones, wherein the planets clear the disk from the inside out, and ``disk-dominated" cases, in which the disk clears the planets, their ultimate fate depending on what happens at the star-disk interface (not resolved in our simulations).  The system depicted in Fig. 1 represents an intermediate case, with more examples shown in the Supporting Material.  Whether planets accumulate at the original inner disk edge \cite{1996Natur.380..606L,2002ApJ...574L..87K,2007ApJ...654.1110T} or in an inner hole, migration tends to produce crowded systems, leading to the excitation of eccentricities via planet-planet scattering and resonances\cite{2003Icar..163..290A,2007astro.ph..3160J,2007astro.ph..3166C,2008arXiv0801.1926L}.
We do not know where 
the true distribution of disk properties falls, but because 6-7\% of Sun-like stars are observed to harbor giant planets\cite{2005PThPS.158...24M},
at least that fraction of disks must fall into the giant planet-bearing upper-left region of Fig. 2.   Despite the speed of the hybrid code, it is as yet not computationally feasible to perform simulations of a sufficient fidelity and number to allow a detailed comparison to the statistical properties of the discovered exoplanets.  
Instead, we performed a smaller set of lengthier simulations, focusing principally on a part of the planet-bearing region.  These produced qualitative agreement with some of the key features of the exoplanet ensemble, namely the correlation of host-star metallicity with planet occurrence \cite{2003ASPC..294..117F,2004A&A...415.1153S}, the distribution of planet masses, planet orbital periods, and the mass-eccentricity distribution\cite{2007ARA&A..45..397U}, as well as insights into how these features arose (see Supporting Material for details).  

Our results also suggest how the Solar System fits into the picture.  In systems with $\tau_{\rm giant} \sim \tau_{\rm disk}$, gas giants do form, but undergo only modest migration and eccentricity growth; thus, it is here where we would most naturally expect to find a Solar System-like outcome.
Figure 2 shows that these cases occupy a relatively narrow region within the parameter space, roughly a diagonal line extending from $(\alpha=1 \times 10^{-3} ,M_{\rm disk}=0.03 M_{\rm Sun})$ to $(\alpha=10^{-2},M_{\rm disk} = 0.08 M_{\rm Sun})$.  Thus, whatever the true distribution of disks within Fig. 2---unless it just happens that disks with $\tau_{\rm giant} \sim \tau_{\rm disk}$ are somehow preferred\cite{vdisk_science_note1}---it is likely that only a minority will lie within this region.
Furthermore, even within this subset there are large stochastic variations, as evidenced by Fig. 2; in only one of the outcomes $(\alpha=3 \times 10^{-3} ,M_{\rm disk}=0.05 M_{\rm Sun})$ do the gas giants bear a reasonable resemblance to Jupiter and Saturn (for an animated example see Supporting Material).
The other potential pathway to a Solar System analogue are cases in which all gas giants except two are engulfed by the star; however, since both survivors must somehow themselves undergo little migration, such outcomes also appear improbable (Fig. 2 shows no candidates for this scenario).
All of this leads us to predict that within the diverse ensemble of planetary systems, ones resembling our own are the exception rather than the rule.  Observations may be hinting at this already\cite{2004MNRAS.354..763B}, though the true planet distribution remains largely obscured by selection effects\cite{2005PThPS.158...24M}.  On the other hand,  scaled-down versions of the Solar System, in which a moderate amount of migration took place, are likely to be more common; indeed, such a system has recently been discovered via microlensing\cite{2008Sci...319..927G}.  Finally, scenarios in which type II migration is reduced\cite{2007Icar..191..158M,2008arXiv0802.1114I} would modify our prediction, permitting a more common occurrence of Solar System analogues.

In all of our simulations, the formation of a gas giant brings with it violent scattering of neighboring smaller bodies, including other cores about to undergo runaway gas accretion themselves.  Such scattering has been proposed as the origin of Uranus and Neptune\cite{1999Natur.402..635T},
with dynamical friction from the remnant outer planetesimal disk (not modelled here) serving to prevent their ejection and ultimately re-circularize their orbits.   Thus, whether or not Jupiter and Saturn analogues are rare, it is likely that Uranus and Neptune analogues are quite common.


\begin{thebibliography}{10}

\bibitem{2001ApJ...553L.153H}
K.~E. {Haisch}, E.~A. {Lada}, C.~J. {Lada}, {\it \apjl\/} {\bf 553}, L153
  (2001).

\bibitem{2007astro.ph..3160J}
M.~{Juric}, S.~{Tremaine}, {\it ArXiv Astrophysics e-prints\/}  (2007).

\bibitem{2007astro.ph..3166C}
S.~{Chatterjee}, E.~B. {Ford}, F.~A. {Rasio}, {\it ArXiv Astrophysics
  e-prints\/} {\bf astro-ph/0703166} (2007).

\bibitem{2007prpl.conf..685U}
S.~{Udry}, D.~{Fischer}, D.~{Queloz}, {\it Protostars and Planets V\/},
  B.~{Reipurth}, D.~{Jewitt}, K.~{Keil}, eds. (2007), pp. 685--699.

\bibitem{2002ApJ...567..596L}
M.~H. {Lee}, S.~J. {Peale}, {\it \apj\/} {\bf 567}, 596 (2002).

\bibitem{1980ApJ...241..425G}
P.~{Goldreich}, S.~{Tremaine}, {\it \apj\/} {\bf 241}, 425 (1980).

\bibitem{1986ApJ...309..846L}
D.~N.~C. {Lin}, J.~{Papaloizou}, {\it \apj\/} {\bf 309}, 846 (1986).

\bibitem{1997Icar..126..261W}
W.~R. {Ward}, {\it Icarus\/} {\bf 126}, 261 (1997).

\bibitem{2004ApJ...604..388I}
S.~{Ida}, D.~N.~C. {Lin}, {\it \apj\/} {\bf 604}, 388 (2004).

\bibitem{2005A&A...434..343A}
Y.~{Alibert}, C.~{Mordasini}, W.~{Benz}, C.~{Winisdoerffer}, {\it \aap\/} {\bf
  434}, 343 (2005).

\bibitem{2006ApJ...652L.133C}
J.~E. {Chambers}, {\it \apjl\/} {\bf 652}, L133 (2006).

\bibitem{2008ApJ...673..487I}
S.~{Ida}, D.~N.~C. {Lin}, {\it \apj\/} {\bf 673}, 487 (2008).

\bibitem{2003Icar..163..290A}
F.~C. {Adams}, G.~{Laughlin}, {\it Icarus\/} {\bf 163}, 290 (2003).

\bibitem{2008arXiv0801.1926L}
A.~T. {Lee}, E.~W. {Thommes}, F.~A. {Rasio}, {\it ArXiv Astrophysics
  e-prints\/} {\bf astro-ph/0801.1926v1} (2008).

\bibitem{2005ApJ...626.1033T}
E.~W. {Thommes}, {\it \apj\/} {\bf 626}, 1033 (2005).

\bibitem{1998AJ....116.2067D}
M.~J. {Duncan}, H.~F. {Levison}, M.~H. {Lee}, {\it \aj\/} {\bf 116}, 2067
  (1998).

\bibitem{2007ApJ...656L..25T}
E.~W. {Thommes}, L.~{Nilsson}, N.~{Murray}, {\it \apjl\/} {\bf 656}, L25
  (2007).

\bibitem{1999..Murray..Dermott..book}
C.~D. {Murray}, S.~F. {Dermott}, {\it Solar System Dynamics\/} (Cambridge
  University Press, 1999).

\bibitem{2000prpl.conf..401H}
D.~J. {Hollenbach}, H.~W. {Yorke}, D.~{Johnstone}, {\it Protostars and Planets
  IV\/} pp. 401--+ (2000).

\bibitem{1998ApJ...495..385H}
L.~{Hartmann}, N.~{Calvet}, E.~{Gullbring}, P.~{D'Alessio}, {\it \apj\/} {\bf
  495}, 385 (1998).

\bibitem{1973A&A....24..337S}
N.~I. {Shakura}, R.~A. {Syunyaev}, {\it \aap\/} {\bf 24}, 337 (1973).

\bibitem{1996Natur.380..606L}
D.~N.~C. {Lin}, P.~{Bodenheimer}, D.~C. {Richardson}, {\it \nat\/} {\bf 380},
  606 (1996).

\bibitem{2002ApJ...574L..87K}
M.~J. {Kuchner}, M.~{Lecar}, {\it \apjl\/} {\bf 574}, L87 (2002).

\bibitem{2007ApJ...654.1110T}
C.~{Terquem}, J.~C.~B. {Papaloizou}, {\it \apj\/} {\bf 654}, 1110 (2007).

\bibitem{2005PThPS.158...24M}
G.~{Marcy}, {\it et~al.\/}, {\it Progress of Theoretical Physics Supplement\/}
  {\bf 158}, 24 (2005).

\bibitem{2003ASPC..294..117F}
D.~A. {Fischer}, J.~A. {Valenti}, {\it Scientific Frontiers in Research on
  Extrasolar Planets\/}, D.~{Deming}, S.~{Seager}, eds. (2003), vol. 294 of
  {\it Astronomical Society of the Pacific Conference Series\/}, pp. 117--128.

\bibitem{2004A&A...415.1153S}
N.~C. {Santos}, G.~{Israelian}, M.~{Mayor}, {\it \aap\/} {\bf 415}, 1153
  (2004).

\bibitem{2007ARA&A..45..397U}
S.~{Udry}, N.~C. {Santos}, {\it \araa\/} {\bf 45}, 397 (2007).

\bibitem{vdisk_science_note1}
{Scenarios have been proposed in which planets themselves are the source of
  viscosity\cite{2001ApJ...552..793G,2004ApJ...606L..77S}; although this would
  cause the two timescales to be correlated, $\tau_{\rm giant}$ would
  systematically be less than $\tau_{\rm disk}$, thus making significant
  migration likely in all planetary systems} .

\bibitem{2004MNRAS.354..763B}
M.~E. {Beer}, A.~R. {King}, M.~{Livio}, J.~E. {Pringle}, {\it \mnras\/} {\bf
  354}, 763 (2004).

\bibitem{2008Sci...319..927G}
B.~S. {Gaudi}, {\it et~al.\/}, {\it Science\/} {\bf 319}, 927 (2008).

\bibitem{2007Icar..191..158M}
A.~{Morbidelli}, A.~{Crida}, {\it Icarus\/} {\bf 191}, 158 (2007).

\bibitem{2008arXiv0802.1114I}
S.~{Ida}, D.~N.~C. {Lin}, {\it ArXiv e-prints\/} {\bf 802} (2008).

\bibitem{1999Natur.402..635T}
E.~W. {Thommes}, M.~J. {Duncan}, H.~F. {Levison}, {\it \nat\/} {\bf 402}, 635
  (1999).

\bibitem{2001ApJ...552..793G}
J.~{Goodman}, R.~R. {Rafikov}, {\it \apj\/} {\bf 552}, 793 (2001).

\bibitem{2004ApJ...606L..77S}
R.~{Sari}, P.~{Goldreich}, {\it \apjl\/} {\bf 606}, L77 (2004).

\bibitem{1981PThPh..70...35H}
C.~{Hayashi}, {\it Prog.~Theor.~Phys.\/} {\bf 70}, 35 (1981).

\bibitem{1998Icar..131..171K}
E.~{Kokubo}, S.~{Ida}, {\it Icarus\/} {\bf 131}, 171 (1998).

\bibitem{2000ApJ...544..481B}
G.~{Bryden}, D.~N.~C. {Lin}, S.~{Ida}, {\it \apj\/} {\bf 544}, 481 (2000).

\bibitem{2000ApJ...537.1013I}
M.~{Ikoma}, K.~{Nakazawa}, H.~{Emori}, {\it \apj\/} {\bf 537}, 1013 (2000).

\end{thebibliography}

%
%


\bibliographystyle{Science}

\begin{scilastnote}
\item This research was supported by the Spitzer Space Telescope Cycle-4 Theoretical Research Program, by the NSF, and by NSERC.
\end{scilastnote}




%

\newpage
\begin{figure}[h]
\vspace{-2.0in}
\epsfig{file=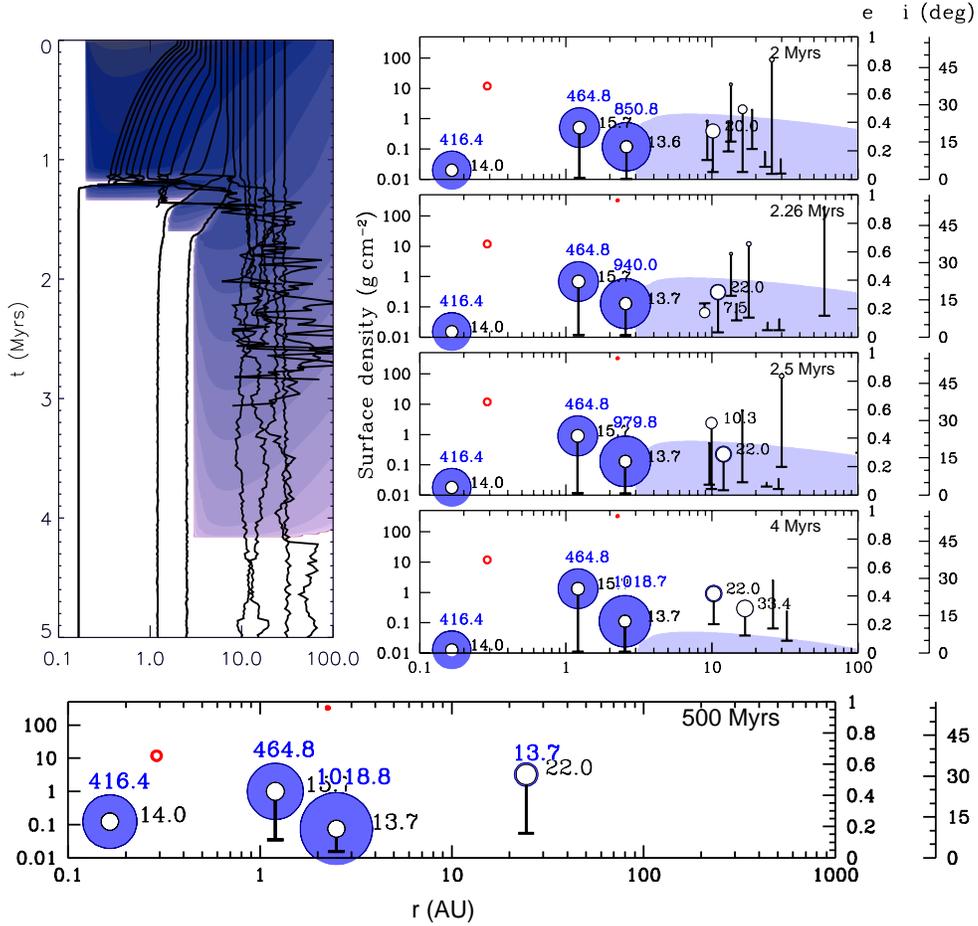, width=6in}
\label{time_evol_1}
\vspace{-2.0in}
\caption{Example of a planet formation simulation.  A Solar-metallicity disk with initial mass $M_{\rm disk}=0.088 M_{\rm Sun}$ and viscosity parameter $\alpha=7.5 \times 10^{-3}$, containing protoplanets of initial mass $10^{-3} M_{\rm E}$ between 2 and 30 AU, evolves for 0.5 Gyr.  Top Left: Planet semi-major axis over time, with the azimuthally-averaged gas disk surface density overlaid as a contour plot.  Right, top four panels:  Snapshots of the system at different intermediate times, showing planet eccentricity (right scale) and inclination (far-right scale, indicated by a horizontal tickmark connected to each planet by a vertical line) versus semi-major axis.  Planet cores (black empty circles) together with whatever gas envelopes they have accreted (dark blue) are labelled with their mass in $M_{\rm E}$ (black and dark blue, respectively).  Planets crossing the inner simulation boundary at 0.1 AU are removed, and their final mass and orbital elements shown (empty red circles).  The azimuthally-averaged disk surface density is also shown (light blue, left scale).  One planet ends up at the inner disk edge (likely $<0.1$ AU if due to the star's magnetospheric cavity\cite{1996Natur.380..606L}; however, we set it at 0.2 AU for computational reasons).
Bottom:  The state of the system at 0.5 Gyrs.  Orbital elements are averages over the last 1 Myr.
}
\end{figure}

\newpage



\begin{figure}[h]
\vspace{-1.5in}
\epsfig{file=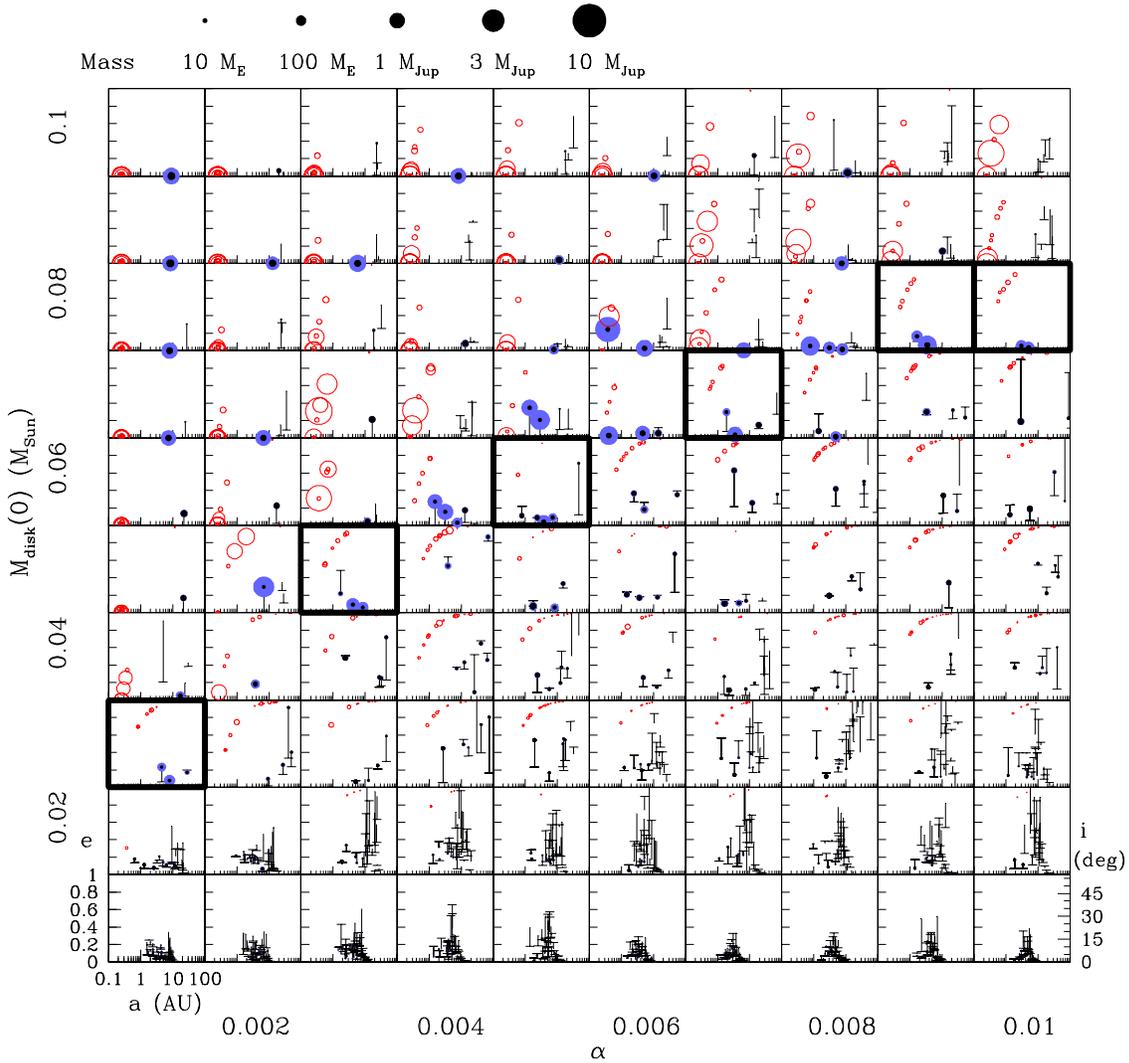, width=6in}
\caption{Final outcomes of a set of 100 simulations, spanning $10^{-3}$ to $10^{-2}$ in viscosity parameter $\alpha$, and $0.01 M_{\rm Sun}$ to $0.1 M_{\rm Sun}$ in initial disk mass.  Simulations are ended after 0.5 Gyrs have elapsed, or if they fail to produce any gas giants within the gas disk's lifetime.  Planet semi-major axes, eccentricities and inclinations (denoted as in Fig. 1) are plotted, as well as the relative solids and gas content of each planet (solid core: black; gaseous envelope: dark blue; size $\propto$ mass$^{1/3}$, see mass scale at top).  To keep computational cost reasonable, simulations have an inner boundary at 0.25 AU, beyond the initial inner edge of the gas disk; any body that crosses the boundary 
is removed, and a red circle is plotted showing its orbital elements and mass at the time of removal.  Toward high $M_{\rm disk}$ and low $\alpha$, planets form early and often during the gas disk's lifetime, most migrate extensively, and many acquire high eccentricities in the process (See Fig. 1 and Supporting Material).  Toward low $M_{\rm disk}$ and high $\alpha$, planet formation is too slow to produce any gas giants during the disk lifetime.  Between these two extremes is a relatively narrow boundary region (thick borders) in which gas giants migrate little and remain at low eccentricity, thus producing some outcomes more similar to the Solar System.  No gas giants at all form in disks of $0.02 M_{\rm Sun}$ or less; for comparison, this is the approximate lower limit on the Solar System's birth disk, called the ``minimum mass Solar nebula"\cite{1981PThPh..70...35H}.}
\end{figure}



\newpage

\begin{figure}[h]
\vspace{-1.5in}
\epsfig{file=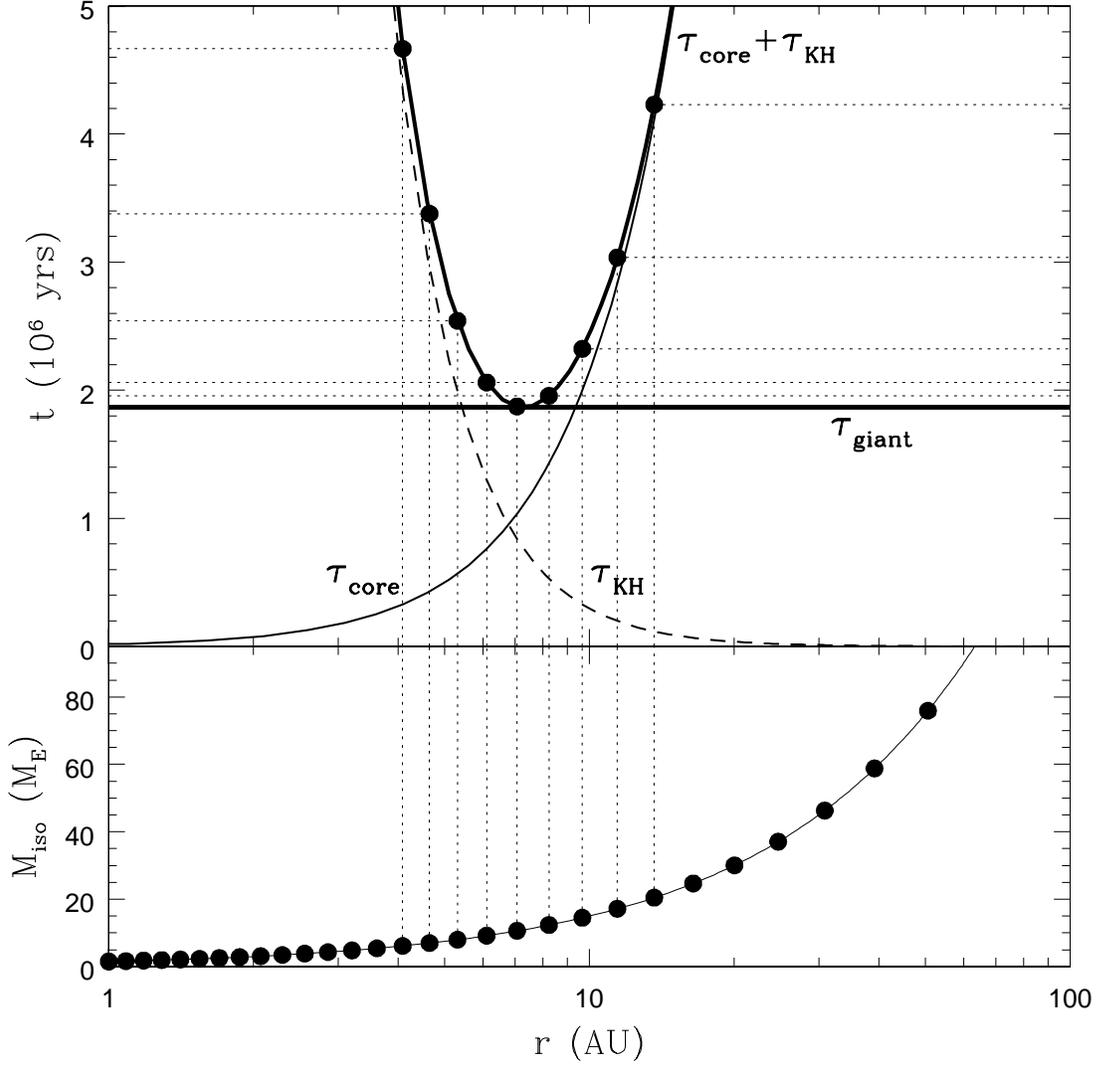, width=6in}
\label{formation time}
\caption{Approximate timing and location of gas giant formation in a protoplanetary disk.  Bottom panel:  The final or ``isolation" mass of solid cores (black dots), with spacing between successive cores taken from planet formation simulations\cite{1998Icar..131..171K}.
Top:  The time (thick solid curve) for a core (black dots; vertical dotted lines connect to corresponding core in bottom panel) to become a gas giant (horizontal dotted lines show times for individual protoplanets).  We approximate this as the sum of the time for the core to reach its final mass, $\tau_{\rm core}$ (thin solid curve), and the time for the core to undergo runaway gas accretion, taken to be its Kelvin-Helmholtz time\cite{2000ApJ...544..481B}, $\tau_{\rm KH}$
(dashed curve). 
As in more detailed calculations\cite{2000ApJ...537.1013I},
we find that gas giant formation commences at one particular radius which for typical parameters lies in or near the Jupiter-Saturn region; in this case at 7 AU, and at time $\tau_{\rm giant}$ just under 2 Myrs).  Giant formation begins in a burst, with several planets growing in rapid succession, then slows down as it spreads to larger and smaller radii.  In practice, once an inner hole forms in the gas disk, formation is constrained to progress only outwards)}
\end{figure}




\end{document}